\documentclass{article}
\usepackage{spconf,amsmath,graphicx}
\usepackage{epsfig,amsmath,amssymb,epsf,amsthm,scalefnt,multirow}
\usepackage{xcolor}
\usepackage{float}
\usepackage{cite}
\usepackage{psfrag}
\usepackage{graphicx}
\usepackage{subcaption}
\newsavebox{\imagebox}

\def\b0{{\pmb{0}}} 

\def\ba{{\mathbf{a}}} \def\bb{{\mathbf{b}}}  
   \def\bh{{\mathbf{h}}}

  \def\bs{{\mathbf{s}}} 
   
\def\by{{\mathbf{y}}} \def\bz{{\mathbf{z}}}

   \def\bH{{\mathbf{H}}}

  \def\bS{{\mathbf{S}}}

\title{Multi-Vehicle Velocity Estimation Using IEEE 802.11ad Waveform}
\name{Geonho Han, Sucheol Kim, and Junil Choi\thanks{This work was partly supported by the National Research Foundation of Korea (NRF) grant funded by the MSIT of the Korea government (No. 2019R1C1C1003638) and by Institute of Information \& communications Technology Planning \& Evaluation (IITP) grant funded by the Korea government (MSIT) (No.2020-0-01882, The Development of Dangerous status recognition Platform in building based on WiFi Low Power Wireless signal sensing without sensor or video camera).}}
\address{School of Electrical Engineering\\
	Korea Advanced Institute of Science and Technology\\
	Emails: \{ghhan6, loehcusmik, junil\}@kaist.ac.kr}
\begin{document}
%
\maketitle
\begin{abstract}
Wireless communication systems are to use millimeter-wave (mmWave) spectra, which can enable extra radar functionalities. In this paper,~we propose a multi-target velocity estimation technique using IEEE 802.11ad waveform in a vehicle-to-vehicle (V2V) scenario. We form a wide beam to consider multiple target vehicles. The Doppler shift of each vehicle is estimated from least square estimation (LSE) using the round-trip delay obtained from the auto-correlation property of Golay complementary sequences in IEEE 802.11ad waveform, and the phase wrapping is compensated by the Doppler shift estimates of proper two frames. Finally, the velocities of target vehicles are obtained from the estimated Doppler shifts. Simulation results show the proposed velocity estimation technique can achieve significantly high accuracy even for short coherent processing interval (CPI).
\end{abstract}
\section{Introduction}
The use of millimeter-wave (mmWave) band for wireless~ communications becomes necessary to achieve extremely high data rates required for fifth-generation (5G) communication systems \cite{Andrews:2014}, which may enable dual functionalities of radar and communications due to the strong directivity of mmWave spectrum. The dual functional systems can be optimized especially for automated vehicles that require both vehicle-to-everything (V2X) and environment sensing systems \cite{Choi:2016,Liu:2020,Hobert:2015}.

There has been much work on integrating radar functionality using wireless communication waveforms. In \cite{Sturm:2009}, radar processing with orthogonal frequency division multiplexing (OFDM) waveform was developed for simultaneous dual operations. In \cite{Li:2019}, a neural network was used to implement a radar system using reflected OFDM signals from multiple targets. The authors in \cite{Duggal:2020} proposed an IEEE 802.11ad-based framework to improve false-alarm rate and the peak-to-sidelobe level for single-target and multi-target scenarios with high Doppler. In \cite{Kumari:2017}, radar functionalities based on IEEE 802.11ad waveform for long range radar were implemented in vehicular environments where multi-target Doppler shifts were estimated from a delay-Doppler map.

In this paper, we propose a novel velocity estimation method using the IEEE 802.11ad waveform for a multi-target vehicle-to-vehicle (V2V) scenario, which outperforms the scheme in \cite{Kumari:2017}. To cover the azimuth~region for the multiple target vehicles, we form a wide beam from the weighted linear combination of several beams. After estimating round-trip delay using the auto-correlation property of Golay complementary sequences embedded in the preamble of IEEE 802.11ad waveform, the Doppler shift can be first recovered by least square estimation (LSE) using echo signals and refined through the phase wrapping compensation. Numerical results demonstrate the proposed technique can estimate the velocities of multiple vehicles accurately within a very short period of time. Our multi-target velocity estimation scheme can be beneficial to improve the performance of localization approaches \cite{Fascista:2020,Kim:2020}.


\begin{figure}
	\centering
	\includegraphics[width=0.554\columnwidth]{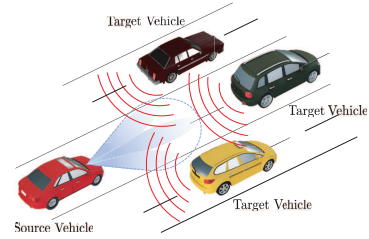}
	\caption{A V2V scenario considered in this paper.}\label{figure1}
\end{figure}

\section{System Model}\label{sec2}
We consider a multi-target scenario where each vehicle is assumed as a single-target model as shown in Fig. \ref{figure1}. A source vehicle transmits the IEEE 802.11ad waveform with a wide beam to estimate Doppler shifts of multiple target vehicles. We first develop a multi-target radar channel model and a wide beam design method and then explain transmit and echo signal models using the IEEE 802.11ad waveform.

\subsection{Radar Channel Model and Beam Design}
The source vehicle is equipped with uniform planar array (UPA) with $N_{\textrm{TX}}=N_{\textrm{TX}}^x\times N_{\textrm{TX}}^y$ for transmit (TX) antennas and $N_{\textrm{RX}}=N_{\textrm{RX}}^x\times N_{\textrm{RX}}^y$ for radar receive (RX) antennas. Assuming that the TX and RX antennas at the source vehicle are co-located, we develop a two-way radar channel model with $P$ target vehicles as \cite{Hong:2014,Jian:2007}
\begin{align}\label{eq1}
\bH(t)=&\sum_{p=0}^{P-1}\sqrt{G_p(t)}\beta_pe^{j2\pi t\nu_p(t)}e^{-j2\pi f_{\textrm{c}}\tau_p(t)}\notag\\
&\qquad\quad\,\,\times\ba_{\textrm{RX}}^*(\phi_p(t),\theta_p(t))\ba_{\textrm{TX}}^{\mathrm{H}}(\phi_p(t),\theta_p(t)),
\end{align}
where $G_p(t)$ is the large scale channel gain caused by path-loss and radar cross section (RCS), $\beta_p\sim\mathcal{CN}(0,1)$ represents the small scale channel gain, $\nu_p(t)=2v_p(t)/\lambda$ is the Doppler shift with the relative velocity $v_p(t)$ and the wavelength $\lambda$ corresponding to carrier frequency $f_{\textrm{c}}$, $\tau_p(t)=2r_p(t)/c$ denotes the round-trip delay with the Euclidean distance from the source vehicle $r_p(t)$ and the speed of light $c$, and $\phi_p(t)$ and $\theta_p(t)$ are the azimuth and elevation angle of arrivals (AoAs) at the source vehicle. In (\ref{eq1}), $\ba_{\textrm{RX}}(\phi,\theta)$ and $\ba_{\textrm{TX}}(\phi,\theta)$ are the UPA steering vectors at the source vehicle.

To detect multiple target vehicles, the TX and RX beams at the source vehicle need to cover wide azimuth angle region of interest. To generate a wide beam~satisfying specific 3 dB azimuth beamwidth, we develop the TX beamformer as \cite{Lee:2019}
\begin{align}\label{eq2}
\mathbf{f}_{\textrm{TX},x}=\sum_{i=1}^{N_{\textrm{c}}}\gamma_i&\ba_{\textrm{TX},x}(\varphi_i,\vartheta_c),\quad \mathbf{f}_{\textrm{TX},y}=\ba_{\textrm{TX},y}(\vartheta_c),\notag\\
\mathbf{f}_{\textrm{TX}}&=\frac{\mathbf{f}_{\textrm{TX},x}\otimes\mathbf{f}_{\textrm{TX},y}}{\|\mathbf{f}_{\textrm{TX},x}\otimes\mathbf{f}_{\textrm{TX},y}\|},
\end{align}
where $N_{\textrm{c}}$ is the number of combined beams, $\varphi_i$ denotes the azimuth angle of the $i$-th beam, $\vartheta_c$ is the fixed elevation angle of the beam, $\gamma_i$ is the weight coefficient for the $i$-th beam, $\otimes$ is the Kronecker product, and $\ba_{\textrm{TX},x}(\varphi,\vartheta)$ and $\ba_{\textrm{TX},y}(\vartheta)$ represent the TX steering vectors for $x$- and $y$-axes, which are given as
\begin{align}\label{eq3}
\ba_{\textrm{TX},x}(\varphi,\vartheta)&=\begin{bmatrix}1,e^{j\psi_x},\cdots,e^{j(N_{\textrm{TX}}^x-1)\psi_x}\end{bmatrix}^{\textrm{T}},\notag\\
\ba_{\textrm{TX},y}(\vartheta)&=\begin{bmatrix}1,e^{j\psi_y},\cdots,e^{j(N_{\textrm{TX}}^y-1)\psi_y}\end{bmatrix}^{\textrm{T}}.
\end{align}
In (\ref{eq3}),
$\psi_x=2\pi d_x\mathrm{cos}(\vartheta)\mathrm{sin}(\varphi)/\lambda$ and $ \psi_y=2\pi d_y\mathrm{sin}(\vartheta)/\lambda$ with the antenna spacing $d_x=d_y=\lambda/2$.

For the radar RX beamformer, the source vehicle can use the conjugate of TX beamformer conveyed from the communication module as $\mathbf{f}_{\textrm{RX}}=\mathbf{f}_{\textrm{TX}}^*$ by the channel reciprocity as in the time division duplexing (TDD) systems for sufficiently short coherent processing interval (CPI) \cite{Chen:2019}. We assume $\mathbf{f}_{\textrm{TX}}$ and $\mathbf{f}_{\textrm{RX}}$ are fixed during one CPI where the RX steering vectors can be defined as in (\ref{eq3}).

\subsection{Transmit and Echo Signal Models}
The transmitted signal at the source vehicle using the IEEE 802.11ad waveform is represented as
\begin{equation}\label{eq4}
x(t)=\sqrt{P_{\textrm{TX}}}\sum_{n=-\infty}^{\infty}s[n]g_{\textrm{TX}}(t-nT_{\textrm{s}}),
\end{equation}
where $P_{\textrm{TX}}$ is the TX power at the source vehicle, $s[n]$ denotes the transmitted symbol with the unit energy $\mathbb{E}\begin{bmatrix}|s[n]|^2\end{bmatrix}=1$, $g_{\textrm{TX}}(t)$ is the TX pulse shaping filter, and $T_{\textrm{s}}$ is the symbol period.

We need to process the echo signal to estimate the vehicular velocities from the Doppler shifts. With the radar channel model in (\ref{eq1}), the echo signal reflected from $P$ target vehicles is given as
\begin{equation}\label{eq5}
y(t)=\sum_{p=0}^{P-1}\sqrt{P_{\textrm{TX}}}h_p(t) x_g(t-\tau_p(t))e^{j2\pi t\nu_p(t)}+z(t),
\end{equation}
where the signal after TX and RX filtering is $x_g(t)=\sum_{n=-\infty}^{\infty}s[n]g(t-nT_{\textrm{s}})$ with the linear convolution of filters $g(t)=g_{\textrm{TX}}(t)*g_{\textrm{RX}}(t)$ that satisfies the Nyquist criterion, $g_{\textrm{RX}}(t)$ is the RX matched filter at the radar module of source vehicle having the same roll-off factor with $g_{\textrm{TX}}(t)$, and $z(t)\sim\mathcal{CN}(0,\sigma_{\textrm{cn}}^2)$ represents the combined term of~clutter and noise. In (\ref{eq5}), $h_p(t)$ denotes the backscattering coefficient determining effective radar channel magnitude. We assume it is constant as $h_p\approx\sqrt{G_p}\beta_p \mathbf{f}_{\textrm{RX}}^{\mathrm{H}}\ba_{\textrm{RX}}^*(\phi_p,\theta_p)\ba_{\textrm{TX}}^{\mathrm{H}}(\phi_p,\theta_p)\mathbf{f}_{\textrm{TX}}$ since the change of time-varying parameters, i.e., $G_p(t), \phi_p(t)$, and $\theta_p(t)$, are negligible for one CPI.

After sampling, the discrete-time representation of echo signal is given as
\begin{equation}\label{eq6}
y[m,k]=\sum_{p=0}^{P-1}\sqrt{P_{\textrm{TX}}}h_pe^{j2\pi \nu_p^m(k+mK)T_{\textrm{s}}}s[k-\ell_p^m]+z[m,k],
\end{equation}
where $m=0,1,\cdots,M-1$ and $k=\ell_0^m,\ell_0^m+1,\cdots,K_{\textrm{pre}}-1+\ell_{0}^m$ are the frame and sample indices with the number of frames $M$ in one CPI and the number of transmitted training samples $K$ during one frame, $K_{\textrm{pre}}$ is the number of preamble samples, and $\nu_p^m$ and $\ell_p^m$ denote the Doppler shift and round-trip delay of the $p$-th target vehicle at the $m$-th frame considering $\ell_{\alpha}^{m}<\ell_{\beta}^{m}$ for $\alpha<\beta$ without loss of generality.

\section{Delay Estimation}
We employ the preamble of IEEE 802.11ad waveform with Golay complementary sequences, which are $\ba_{128}$ and $\mathbf{b}_{128}$, having auto-correlation property useful for target sensing \cite{Muns:2019}. The cross-correlation between the echo signal and a specific part of preamble, denoted as $\bs_{\textrm{c}}$, starting from 2049-th sample\footnote{We refer to \cite{Han:2020GLOBE} for the reason of using this specific part of preamble.} can be defined as
\begin{align}\label{eqd1}
R_{\bs_{\textrm{c}}\tilde{\by}_m}[\ell]&=\sum_{k_c=0}^{511}s_{\textrm{c}}[k_c]\tilde{y}_m^*[\ell+k_c]\notag\\
&=\sum_{k_c=0}^{511}\sum_{p=0}^{P-1}\sqrt{P_{\textrm{TX}}}h_p^*e^{-j2\pi \nu_p^m(\ell+k_c+2048+mK)T_{\textrm{s}}}\notag\\
&\qquad\,\times s_{\textrm{c}}[k_c]s[\ell+k_c+2048-\ell_p^m]+\tilde{\bz}_m^{\mathrm{H}}\bs_{\textrm{c}},
\end{align}
where $\bs_{\textrm{c}}=\begin{bmatrix}-\ba_{128}^{\mathrm{T}}-\bb_{128}^{\mathrm{T}}-\ba_{128}^{\mathrm{T}}\bb_{128}^{\mathrm{T}}\end{bmatrix}^{\mathrm{T}}$, and $\tilde{y}_m[k]=y[m,k+2048]$ and $\tilde{z}_m[k]=z[m,k+2048]$ are the $k$-th elements of $\tilde{\by}_m$ and $\tilde{\bz}_m$ with length 512.

The delay of dominant target vehicle can be estimated as
\begin{equation}\label{eqd2}
\hat{\ell}_{\textrm{D}}^m=\textrm{argmax}_{\ell}\,\,|R_{\bs_{\textrm{c}}\tilde{\by}_m}[\ell]|,
\end{equation}
and the delays of other target vehicles can be estimated by searching $\ell$ around $\hat{\ell}_{\textrm{D}}^m$ leading to $|R_{\bs_{\textrm{c}}\tilde{\by}_m}[\ell]|$ larger than a threshold since it is reasonable to assume the target vehicles are not too far apart from each other. The threshold used in the delay estimation is an upper bound of $\tilde{\bz}_m^{\mathrm{H}}\bs_{\textrm{c}}$ in (\ref{eqd1}) obtained from the Cauchy-Schwarz inequality. The set of estimated delays at the $m$-th frame is $\left\{ \hat{\ell}_0^m,\hat{\ell}_1^m,\cdots,\hat{\ell}_{\textrm{D}}^m,\cdots,\hat{\ell}_{\hat{P}-2}^m,\hat{\ell}_{\hat{P}-1}^m\right\}$ with $\hat{P}$ detected target vehicles, where we assume the target vehicles are perfectly detected, i.e., $\hat{P}=P$.

\section{Doppler Shift Estimation}
To estimate the Doppler shift $\nu_p^m$ from the echo signals, we need to figure out the exponential term $e^{j2\pi \nu_p^m(k+mK)T_{\textrm{s}}}$ in (\ref{eq6}), which can be obtained from the backscattering coefficient $h_p$ and effective radar channel for an arbitrary $m$-th frame $h_pe^{j2\pi \nu_p^m(k+mK)T_{\textrm{s}}}$. We first consider the approximated $0$-th frame echo signal to observe the backscattering coefficient as
\begin{equation}\label{eq7}
y[0,k]\approx\sum_{p=0}^{\hat{P}-1}\sqrt{P_{\textrm{TX}}}s[k-\hat{\ell}_p^0]h_p+z[0,k].
\end{equation}
The approximation in (\ref{eq7}) comes from $e^{j2\pi\nu_p^0kT_{\textrm{s}}}\approx 1$ due to the short symbol period $T_{\textrm{s}}$. After concatenating all $K_{\textrm{pre}}$ samples to make $\by_0=\sqrt{P_{\textrm{TX}}}\bS_0\bh+\bz_0$, we can recover the backscattering coefficient $h_p$ through the LSE as
\begin{equation}\label{eq8}
\hat{\bh}=\frac{(\bS_0^{\mathrm{H}}\bS_0)^{-1}\bS_0^{\mathrm{H}}\by_0}{\sqrt{P_{\textrm{TX}}}},
\end{equation}
where the $p$-th element of $\bh$ is $h_p$, and the $(x_1,x_2)$-th element of $\bS_0$ is represented as $s[\hat{\ell}_0^0+x_1-1-\hat{\ell}_{x_2-1}^0]$ for $0\le \hat{\ell}_0^0+x_1-1-\hat{\ell}_{x_2-1}^0\le K_{\textrm{pre}}-1$ and zero otherwise.

Now, to extract the exponential term $e^{j2\pi\nu_p^m(k+mK)T_{\textrm{s}}}$, we exploit the echo signal of $m_{\textrm{d}}$-th frame. By fixing $k=(2\hat{\ell}_0^{m_{\textrm{d}}}+K_{\textrm{pre}}-1)/2$ on the exponential term, we can obtain the approximated echo signal as
\begin{align}\label{eq9}
&y[m_{\textrm{d}},k]\approx\sum_{p=0}^{\hat{P}-1}\sqrt{P_{\textrm{TX}}}s[k-\hat{\ell}_p^{m_{\textrm{d}}}]\notag\\
&\times\underbrace{h_pe^{j2\pi\nu_p^{m_{\textrm{d}}}((2\hat{\ell}_0^{m_{\textrm{d}}}+K_{\textrm{pre}}-1)/2+m_{\textrm{d}}K)T_{\textrm{s}}}}_{=h_{m_{\textrm{d}}}[p]}+z[m_{\textrm{d}},k].
\end{align}
Similar to the $0$-th frame, we have $\by_{m_{\textrm{d}}}=\sqrt{P_{\textrm{TX}}}\bS_{m_{\textrm{d}}}\bh_{m_{\textrm{d}}}+\bz_{m_{\textrm{d}}}$ by concatenating all received echo samples where $\bS_{m_{\textrm{d}}}$ is defined similarly with $\bS_0$, and the $p$-th element of $\bh_{m_{\textrm{d}}}$ is $h_{m_{\textrm{d}}}[p]$ in (\ref{eq9}). The estimate of $\bh_{m_{\textrm{d}}}$ via the LSE is given as
\begin{equation}\label{eq10}
\hat{\bh}_{m_{\textrm{d}}}=\frac{(\bS_{m_{\textrm{d}}}^{\mathrm{H}}\bS_{m_{\textrm{d}}})^{-1}\bS_{m_{\textrm{d}}}^{\mathrm{H}}\by_{m_{\textrm{d}}}}{\sqrt{P_{\textrm{TX}}}}.
\end{equation}
Then, the Doppler shift is obtained using (\ref{eq8}) and (\ref{eq10}) as
\begin{equation}\label{eq11}
\hat{\nu}_p^{m_{\textrm{d}}}=\frac{\angle{(\hat{h}_{m_{\textrm{d}}}[p]/\hat{h}[p])}}{2\pi((2\hat{\ell}_0^{m_{\textrm{d}}}+K_{\textrm{pre}}-1)/2+m_{\textrm{d}}K)T_{\textrm{s}}}.
\end{equation}

The Doppler shift estimate in (\ref{eq11}), however, may suffer from phase wrapping since $\angle{(\hat{h}_{m_{\textrm{d}}}[p]/\hat{h}[p])}$ is restricted to $[-\pi,\pi]$. To compensate the phase wrapping for accurate Doppler shift estimation, we consider the magnitude difference of true Doppler shifts of two specific frames as
\begin{align}\label{eq12}
&c_p=|\nu_p^{m_{\textrm{d}}}|-|\nu_p^{m_{\textrm{i}}}|=|\zeta_p^{m_{\textrm{d}}}D_{m_{\textrm{d}}}|-|\zeta_p^{m_{\textrm{i}}}D_{m_{\textrm{i}}}|\notag\\
&=|(2\pi N_p+\zeta_{p,\textrm{wrap}}^{m_{\textrm{d}}})D_{m_{\textrm{d}}}|-|(2\pi N_p+\zeta_{p,\textrm{wrap}}^{m_{\textrm{i}}})D_{m_{\textrm{i}}}|,
\end{align}
where $\zeta_p^{m_{\textrm{d}}}$ is the true unwrapped phase of $\angle{(\hat{h}_{m_{\textrm{d}}}[p]/\hat{h}[p])}$ in (\ref{eq11}) for the $p$-th target vehicle, $\zeta_{p,\textrm{wrap}}^{m_{\textrm{d}}}$ is the residual phase after the wrapping, $D_{m_{\textrm{d}}}$ is the denominator inverse in (\ref{eq11}), and $N_p$ is the number of phase wrappings that is assumed to be the same for both $m_{\textrm{d}}$-th and $m_{\textrm{i}}$-th frames. Using the magnitude difference of Doppler shift estimates $\hat{c}_p=|\hat{\nu}_p^{m_{\textrm{d}}}|-|\hat{\nu}_p^{m_{\textrm{i}}}|$, the number of wrappings $N_p$ can be represented as
\begin{align}\label{eq13}
N_p=\pm\frac{\hat{c}_p}{2\pi (D_{m_{\textrm{i}}}-D_{m_{\textrm{d}}})}\pm\frac{c_p}{2\pi (D_{m_{\textrm{i}}}-D_{m_{\textrm{d}}})}.
\end{align}
The signs in (\ref{eq13}) depends on the signs of $\zeta_p$ and $\zeta_{p,\textrm{wrap}}$ as $(+,+)$ for $(\zeta_p<0,\,\zeta_{p,\textrm{wrap}}>0)$, $(-,-)$ for $(\zeta_p>0,\,\zeta_{p,\textrm{wrap}}<0)$, $(+,-)$ for $(\zeta_p>0,\,\zeta_{p,\textrm{wrap}}>0)$, and $(-,+)$ for $(\zeta_p<0,\,\zeta_{p,\textrm{wrap}}<0)$. Although $c_p$ is unknown in practice, the role of $c_p$ is just to make $N_p$ be an integer. Therefore, we can approximate $N_p$ as
\begin{align}\label{eq14}
\bar{N}_p\approx\begin{cases}\lfloor\frac{\hat{c}_p}{2\pi (D_{m_{\textrm{i}}}-D_{m_{\textrm{d}}})}\rceil,&\textrm{for}\quad\zeta_{p,\textrm{wrap}}>0\\
\lfloor-\frac{\hat{c}_p}{2\pi (D_{m_{\textrm{i}}}-D_{m_{\textrm{d}}})}\rceil,&\textrm{for}\quad\zeta_{p,\textrm{wrap}}<0
\end{cases},
\end{align}
where $\lfloor\cdot\rceil$ rounds the argument into the closest integer. The Doppler shift is finally refined using $\bar{N}_p$ as $\hat{\hat{\nu}}_p^{m_{\textrm{d}}}={\hat{\nu}}_p^{m_{\textrm{d}}}+2\pi\bar{N}_pD_{m_{\textrm{d}}}$.

The velocities of target vehicles $V_p$ are obtained considering the small azimuth angle between the vehicles on the road~as
\begin{align}\label{eq15}
\nu_p^{m_{\textrm{d}}}=\frac{2v_{p}^{m_{\textrm{d}}}}{\lambda}=&\frac{2(V_{\textrm{s}}-V_p)\mathrm{cos}(\phi_p^{m_{\textrm{d}}})}{\lambda}\approx\frac{2(V_{\textrm{s}}-V_p)}{\lambda},\notag\\
&\hat{V}_p\approx V_{\textrm{s}}-\frac{\hat{\hat{\nu}}_p^{m_{\textrm{d}}}\lambda}{2},
\end{align}
where $v_{p}^{m_{\textrm{d}}}$ and $\phi_p^{m_{\textrm{d}}}$ are the relative velocity and azimuth angle of the $p$-th target vehicle at the $m_{\textrm{d}}$-th frame, and $V_{\textrm{s}}$ is the velocity of source vehicle.

\begin{figure}
	\centering
	\includegraphics[width=0.71\columnwidth]{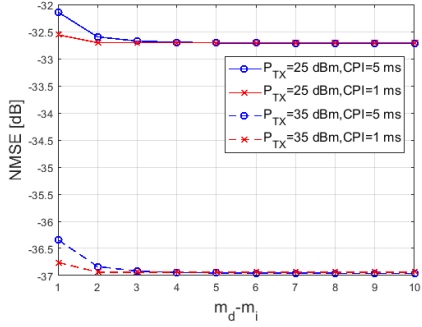}
	\caption{NMSE versus $m_{\textrm{d}}-m_{\textrm{i}}$ of proposed technique.}\label{figure2}
\end{figure}

\section{Simulation Results}
We verify the effectiveness of proposed velocity estimation technique through simulation results. The TX and RX antennas of source vehicle are $8\times 2$ UPAs, and the TX pulse shaping filter and RX matched filter are root-raised cosine filters with the roll-off factor 0.25. The three beams are combined to satisfy the 3 dB azimuth beamwidth of 0.4084 rad considering the width of realistic road, and the 3 dB elevation beamwidth is 1.0399 rad. The three target vehicles are considered in the simulations, and the velocities of source and target vehicles are $V_{\textrm{s}}=$ 25.271 m/s, $(V_1,V_2,V_3)=$ (20.279 m/s, 24.949 m/s, 21.806 m/s). We assume 20 dBsm RCS for each target vehicle and the frame in (\ref{eq9}) as $m_{\textrm{d}}=M-1$. The parameters related to IEEE 802.11ad waveform are the carrier frequency $f_{\textrm{c}}=$ 60 GHz, the bandwidth $W=$ 1.76 GHz, the number of training samples in one frame $K_{\textrm{pre}}=$ 3328, the time duration of one frame $T_{\textrm{f}}=KT_{\textrm{s}}$ with $T_{\textrm{s}}=1/W$ and $K=$ 13632, the number of frames in one CPI $M=\lfloor \textrm{CPI}/T_{\textrm{f}} \rfloor$ \cite{Spec:2012}. The variance of $z(t)$ in (\ref{eq5}) is $\sigma_{\textrm{cn}}^2=N_0W+P_c$ with the noise spectral density $N_0=$ -174 dBm/Hz and the clutter power $P_c$ depending on $P_{\textrm{TX}}$ \cite{Lacomme:2001}. The threshold used in the delay estimation is $512\sigma_{\textrm{cn}}$. As a performance metric, we rely on the normalized mean square error (NMSE) defined as
\begin{equation}\label{eq16}
\textrm{NMSE}=\frac{1}{P}\sum_{p=0}^{P-1}\mathbb{E}\begin{bmatrix}\left(\frac{V_p-\hat{V}_p}{V_p}\right)^2\end{bmatrix}.
\end{equation}

\begin{figure}
	\centering
	\includegraphics[width=0.71\columnwidth]{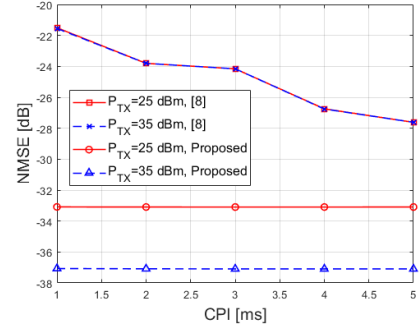}
	\caption{NMSEs according to CPI duration. The proposed technique is based on $m_{\textrm{i}}=m_{\textrm{d}}-6$.}\label{figure3}
\end{figure}

In Fig. \ref{figure2}, the effect of $m_{\textrm{d}}-m_{\textrm{i}}$ in the Doppler shift estimation is verified. If $m_{\textrm{d}}-m_{\textrm{i}}$ is large, the estimation error is reduced for both CPI values. Therefore, it is preferred to pick relatively large $m_{\textrm{d}}-m_{\textrm{i}}$ as long as both frames experience the same number of wrappings. For small $m_{\textrm{d}}-m_{\textrm{i}}$, the shorter CPI case gives better velocity estimate since the phase wrapping happens frequently with long CPI, i.e., with large $M$. If $M$ is large, the round-off error in (\ref{eq14}) could be non-negligible, leading to large NMSE.

In Fig. \ref{figure3}, we compare the proposed estimation method to the scheme in \cite{Kumari:2017}. The proposed technique outperforms the scheme in \cite{Kumari:2017} for all CPI values even with lower TX power than \cite{Kumari:2017}. The gap is excessive especially when CPI is small, which is usually the case when vehicles move fast. The NMSE of scheme in \cite{Kumari:2017} does not vary with different TX power but decreases with CPI since the Doppler resolution is a function of CPI duration. On the contrary, the proposed technique can fully exploit higher TX power because it numerically extracts accurate Doppler shifts from echo signals rather than improving the Doppler resolution with long CPI. This clearly shows the advantage of proposed velocity estimation technique for vehicular environments.

\section{Conclusion}
In this paper, we proposed the multi-target velocity estimation technique based on the IEEE 802.11ad waveform in a V2V scenario. To detect multiple target vehicles, we employed a wide beam designed by the weighted linear combination of beams to cover the azimuth region of interest. The round-trip delay was estimated using the good auto-correlation property of Golay complementary sequences embedded in the preamble of IEEE 802.11ad waveform. The Doppler shift was obtained from the LSE using the echo signals and then refined to compensate the phase wrapping. We verified the high velocity estimation accuracy of proposed technique even within short CPI via simulations.


\bibliographystyle{IEEEtran}
\bibliography{refs_all}

\end{document}